# Basic principle of superconductivity

Tian De Cao

The basic principle of superconductivity is suggested in this paper. There have been two vital wrong suggestions on the basic principle, one is the relation between superconductivity and the Bose-Einstein condensation (BEC), and another is the relation between superconductivity and pseudogap.

*Department of physics, Nanjing University of Information Science & Technology, Nanjing 210044, China*

*to whom correspondence should be addressed. E-mail: tdcao@nuist.edu.cn

The focus of superconductivity is the basic principle of superconductivity, while some suggestions missed the discovery of this principle. One of suggestions is the BEC-based superconductivity [1, 2]. The key problem related to the origin is the pseudogap, while there are some misunderstandings to the results of experiments. One of examples is the Kondo's conclusion [3] that the pseudogap competes with the superconductivity.

We find that superconductivity is not from the BEC of pairs, and there is the direct evidence of preformed pairs.

In a superconductor, a Cooper pair ($\vec{k}\uparrow, -\vec{k}\downarrow$) is the name given to electrons that are bound together at the temperature $T \leq T^*$ with opposite momentum and opposite spin first described in 1956 by Cooper [4]. The Cooper pair has zero spin and is called the singlet pair. The triplet pair ($\vec{k}\uparrow, -\vec{k}\uparrow$) is two electrons bound with opposite momentum but parallel spin. The pairs are responsible for superconductivity at the temperature $T \leq T_c$, as described in the BCS theory [5]. The critical temperature $T_c$ is equal to the pairing temperature $T^*$ in normal superconductor. A pair looks like a boson as its total spin is integer (0 or 1), thus some physicists suggested that superconductivity is from the Bose-Einstein condensation (BEC) of pairs, this is the so-called BEC-based viewpoint. This viewpoint argues that Cooper pairs are so many that both the BEC and the pairing occur simultaneously when the temperature is lowed to the pairing temperature $T^*$. Based on the extrapolation of the number density of pairs, the BEC temperature could be $T_c \geq T^*$ for normal superconductors, but the BEC after the pairing results in $T_c = T^*$. The BEC viewpoint also argues that the number density of pairs may be so low that $T_c \leq T^*$ in cuprate superconductors, and this seems an explanation of pseudogap.

However, we find that superconductivity is not from the BEC of pairs.

Firstly, if the number density of Cooper pairs is large enough, the BEC viewpoint should predict that a weak magnetic field could not affect the critical temperature, while experiments show that magnetic field certainly decreases the critical temperature.

Secondly, if a superconductor has so few pairs that only one Cooper pair is in the lowest energy state at $T_c$ on the basis of the BEC viewpoint, does superconductivity occur? As far as we know, it is hard to image that one pair could lead to superconductivity.

Thirdly, if superconductivity was from the BEC of pairs, because these pairs, whether free or interacting, should obey the Bose-Einstein distribution, some pairs must have various momentums $\vec{q}$, thus the pairing functions should be similar to the form $<c_{-\vec{k}\downarrow}c_{\vec{k}+\vec{q}\uparrow}>$, various pairs should have different bound energies, and thus it is hard to imagine that various pairs are formed in the same pairing temperature $T^*$. That is to say, the pseudogap temperature could not be well defined on the basis of the BEC viewpoint.

Thus we conclude that superconductivity is not from the BEC.

The electron pairs do not obey the Bose-Einstein distribution, this is because one pair is replaced unceasingly by another pair, and the lifetime of a pair is short, while a pair has the mass of two electrons.

Then, how does superconductivity occur? A possible basic principle is that various superconductivities originate from the electron pairing around the Fermi surface and there may exist the pseudogap state associated with preformed pairs far from the Fermi surface.

Some questions may arise.

Firstly, why is superconductivity from pairs behaving as bosons? Because a state could be occupied by many bosons, many pairs around the Fermi surface could be in the same state which has zero momentum (or same momentum for forming super-current), and these pairs contribute to superconductivity.

Secondly, are there the evidences of performed pairs? Yes, there are positive results [6]. Here we comment another "counterexample" [3]. Kondo and his coauthors conclude that the pseudogap competes with the high temperature superconductivity in the cuprates on the basis of an angle resolved photoemission spectroscopy (ARPES), while we find that their observation is just the evidence of preformed pairs.

They perform a straight forward quantitative analysis of the energy distribution curves (EDC) from the ARPES experiment where the symmetrized EDC may directly describe the spectral function of electron systems [7], and they investigated the doping, temperature and momentum dependence of the coherent spectral weight and pseugodap one in Bi2201 samples. They find that there is a direct correlation between the suppression in the low energy spectral weight $W_{PG}$ due to the pseudogap (it should be the superconducting gap for $T \leq T_c$ with our viewpoint) and the increase in the coherent spectral weight $W_{CP}$ due to the paired electrons for all factors dependent.

To understand the Kondo's ARPES experiments on the basis of the basic principle, one could note that some electrons begin to be paired in the antinodal region far from the Fermi surface when the temperature arrives at $T^*$ for the Bi2201 samples at a certain doping. The number of preformed pairs increases with the decreased temperature, and the pairing space will be close to the Fermi surface (such as the antinodal region of the Brillouin zone) when the temperature is lowed toward $T_c$. As soon as the temperature arrives at $T \leq T_c$, the electron pairs appear around the Fermi surface, and this pseudogap becomes the superconducting gap.

Because the number of these pairs increases with the decreased temperature and the excitations from the pairing space are suppressed, $W_{PG}$ should increase with the decreased temperature for $T^* > T > T_c$ as shown in ARPES. When the pairing space moves toward around the Fermi surface for $T < T_c$, the increase of the number of pairs can be neglected, $W_{PG}$ should decrease while $W_{CP}$ increase, thus the almost perfect linear anti-correlation between $W_{CP}$ and $W_{PG}$ can be understood for all factors dependent.

Of course, except the pseudogap associated with performed pairs, there may be other pseudogaps which are associated with the spin density wave and the charge density wave. These problems have to be investigated.

**References**

[1] J. M. Singer, M. H. Pedersen, T. Schneider, H. Beck & H.-G. Matuttis, From BCS-like superconductivity to condensation of local pairs: A numerical study of the attractive Hubbard model, Phys. Rev B. 54, 1996, 1286-1301.

[2] A. S. Alexandrov, Theory of giant and nil proximity effects in cuprate semiconductors, Phys. Rev. B 75, 2007, 132501-132505.

[3] T. Kondo, R. Khasanov, T. Takeuchi, J. Schmalian, & A. Kaminski, Direct evidence for competition between the pseudogap and high temperature superconductivity in the cuprates, Nature 457, 2009, 296-300.

[4] L. N. Cooper, Bound electron pairs in a degenerate Fermi gas, Physical Review 104 (4), 1956, 1189–1190.

[5] J. Bardeen, L. N. Cooper, & J. R. Schrieffer, Theory of Superconductivity, Phys. Rev. 108, 1957, 1175-1204.

[6] J. Lee, K. Fujita, A. R. Schmidt, Chung Koo Kim, H. Eisaki, S. Uchida, & J. C. Davis, Spectroscopic Fingerprint of Phase-Incoherent Superconductivity in the Underdoped $Bi_2Sr_2CaCu_2O_{8+\delta}$, Science 325, 2009, 1099-1103.

[7] M. R. Norman, M. Randeria, H. Ding, & J. C. Campuzano, Phenomenology of photoemission lineshapes of high Tc superconductors, 1998, Phys. Rev. B **57**, R11093- R11093.